\documentstyle[12pt]{article}

\begin{document}
\newcommand{\hs}{\hspace{5mm}}
\newcommand{\vs}{\vspace{2mm}}
\newcommand{\alphap}{{\alpha}^{\prime}}
\newcommand{\nup}{{\nu}^{\prime}}
\newcommand{\alphaz}{{\alpha}^{\prime \prime}}
\newcommand{\nuz}{{\nu}^{\prime \prime}}
\newcommand{\z}{\prime \prime}
\newcommand{\p}{\prime}
\begin{titlepage}
\vspace{-10mm}
\begin{flushright}

\end{flushright}
\vspace{12pt}
\begin{center}
\begin{large}
{\bf On the construction of the Hadamard sates in two dimensions  }\\
\end{large}
\vspace{1cm} ${\bf P.~
Moyassari^{\dag}}\footnote{e-mail:~p-moyassari@cc.sbu.ac.ir.} $

 {\small {\dag Department of Physics,
Shahid Beheshti University, Evin,
Tehran 19839,  Iran.}}\\
\end{center}
\abstract{The two dimensional analog of the Hadamard state
condition is used to specify the local Hadamard states associated
with a linear quantum field coupled to a two dimensional
gravitational background. To characterize a local Hadamard state
corresponding to a physical vacuum state, we apply a
superselection rule in which the state dependent part of the
two-point function is determined in terms of a dynamical scalar
field. It implies a basic connection between the vacuum state and
a scalar field coupled to gravity. We study the characteristics of
the Hadamard vacuum state through this superselection rule using
two different background metrics, the two dimensional analog of
the schwarzschild and FRW metric.}
\end{titlepage}
\section{Introduction}
The Green's function or two-point function, is an important
quantity in the study of quantum field theory in curved or flat
space time. In a linear theory the antisymmetric part of the
two-point function (commutator function) is common to all states
in the same representation. Thus, the characteristics of a state
are encoded in the symmetric part of the two-point function
 which denoted in the following by $ G^+(x,x')$. There are some basic
problems concerning inherent ambiguities in the definition of a
physical state associated with a quantum field. In the flat space
time it is always possible to use the Poincare symmetries to
obtain the physical vacuum and the physically admissible states
most naturally arise as local excitations of this state. In the
curved space time this procedure dose not apply, because on a
general curved space time one may not find a global symmetry. In
this case the problem concerning the determination of the local
states and the role of the global features of space time is of
obvious importance. The Hadamard formalism provides a framework in
which we may improve our understanding in this context. This
formalism assumes that the singular part of $ G^+(x,x')$ is given
by the geometrical Hadamard ansatz \cite{1}. In this prescription,
 however, there exist problems in the specification of the state
 dependent part of the two-point function. For characterizing the
 physical states, it therefore seems to be essential to find out a
 suitable scheme for the treatment of these problems. In the present paper
we shall consider this issue for a quantum scalar field coupled to
a two dimensional gravitational background. The organization of
this paper is as follows: In section 2, we present the Hadamard
prescription and briefly review the derivation of the local
constraints on the state dependent part of the two-point function.
In section 3, we develop a dynamical model in order to analyze
these constraints. To characterize a physical vacuum state, we
apply a superselection rule in which the state dependent part of
the two-point function is related to a dynamical scalar field.
This superselection rule implies a basic connection between the
vacuum state and a scalar field coupled to gravity. In sections 4
and 5, we apply this superselection rule using two different
background metrics and investigate the characteristics of the
Hadamard vacuum state. These considerations shall provide the
Hawking radiation at space-like infinity, in the case of the
Schwarzschild metric; and a thermal radiation at the present
epoch, in the context of cosmology.

\section{Hadamard state condition in two dimensions}
We consider a free massless quantum scalar field $\phi(x)$
propagating in a two dimensional gravitational background with the
action functional (In following the semicolon and $\nabla$ denote
covariant differentiation)
\begin{equation}\label{1}
    S[\phi]=-\frac{1}{2}\int
    d^2x\sqrt{-g}\nabla_\mu\phi\nabla^\mu\phi.
\end{equation}
This action gives rise to the field equation
\begin{equation}\label{2}
\Box\phi(x)=0.
\end{equation}
A state of $\phi(x)$ is characterized by a hierarchy of Wightman
functions
\begin{equation}\label{3}
<\phi(x_1),...,\phi(x_n)>.
\end{equation}
We are primarily interested in those states which reflect the
intuitive notion of a vacuum. For this aim, we may restrict
ourselves basically to quasi-free states, i.e., states for which
truncated n-point functions vanish for $n>2$. Such states may be
characterized by their two-point functions. Equivalence principle
suggests that the leading singularity of $G^+(x,x')$, symmetric
part of the two-point function, should have a close correspondence
to the singularity structure of the two-point function of a free
massless field in a two dimensional Minkowski space time. One may
therefore assume that for $x$ sufficiently close to $x'$ the
function $G^+(x,x')$ can be written as \cite{2}
\begin{equation}\label{4}
G^+(x,x')=-\frac{1}{4\pi}\ln\sigma(x,x')+F(x,x'),
\end{equation}
where $\sigma(x,x')$ is one-half the square of the geodesic
distance between $x$ and $x'$ and $F(x,x')$ is a regular function.
This may be viewed as a two dimensional analog of the Hadamard
ansatz \cite{1}. The function $F(x,x')$ satisfies a general
constraint obtained from the symmetry condition of $G^+(x,x')$ and
the requirement that the expression (\ref{4}) satisfies the wave
equation (\ref{2}). The study of this constraint has obviously a
particular significance for analyzing the state dependent part of
the two-point function. It was shown in details in \cite{2} that
the two independent constraints imposed on the state dependent
part of the two-point function have the form
\begin{equation}\label{4a}
    F^\alpha_{\hspace{0.2cm}\alpha}(x)=-\frac{1}{12\pi}R,
\end{equation}
\begin{equation}\label{4b}
F_{\alpha\beta}^{\hspace{0.4cm};\alpha}(x)-\frac{1}{2}F^\alpha_{\hspace{0.2cm}\alpha;\beta}(x)+\frac{1}{12}
    (\Box
    F(x))_{;\beta}-\frac{1}{3}\Box(F_{;\beta}(x))-\frac{1}{12}RF_{;\beta}(x)=\frac{1}{48\pi}R_{;\beta}.
\end{equation}
The functions $F(x)$ and $F_{\alpha\beta}(x)$ are the coefficients
in the covariant expansion of $F(x,x')$, namely
\begin{equation}
\begin{array}{ll}\label{7} \vspace{0.5cm}
F(x,x')=F(x)-\frac{1}{2}F_{;\alpha}\sigma^{;\alpha}(x)+\frac{1}{2}
F_{\alpha\beta}\sigma^{;\alpha}\sigma^{;\beta}(x)\\
\hspace{1.5cm}+\frac{1}{4}
[\frac{1}{6}F_{;\alpha\beta\gamma}(x)-F_{\alpha\beta;\gamma}(x)]\sigma^{;\alpha}\sigma^{;\beta}\sigma^{;\gamma}
+{\mathcal{O}}(\sigma^2).
\end{array}
\end{equation}
It should be noted that in the derivation of constraints
(\ref{4a}) and (\ref{4b}) the covariant expansion of $F(x,x')$ has
been used only up to the second order expansion terms. In general
there exist additional constraints on the higher order terms. By
combining (5) and (6) we establish another relation which can be
written as a total divergence \cite{2}
\begin{equation}\label{5}
    \nabla^\alpha\Sigma_ {\alpha\beta}=0,
\end{equation}
where
\begin{equation}\label{6}
    \Sigma_{\alpha\beta}=\frac{1}{2}(F_{;\alpha\beta}(x)-\frac{1}{2}g_{\alpha\beta}\Box
    F(x))
    -(F_{\alpha\beta}(x)+\frac{1}{48\pi}g_{\alpha\beta}R).
\end{equation}
It is obvious that
\begin{equation}\label{5a}
    \Sigma^{\alpha}_\alpha=\frac{1}{24\pi}R.
\end{equation}
We can consider (\ref{5}) and (\ref{5a}) as the constraints
imposed on the state dependent part of the two-point function. The
function $F(x)$ may be considered as arbitrary and its
specification depends significantly on the choice of a state. Once
a specific assumption has been made on the form of $F(x)$, the
function $F_{\alpha\beta}(x)$ (and hence $G^+$) can completely
determined by constraints (\ref{5}) and (\ref{5a}). It should be
remarked that in general there is a missing length scale in the
expression (\ref{4}) emerging from the fact that the argument of
the logarithm must be dimensionless. Thus, $F(x,x')$ must supply a
term that is the logarithm of a length \cite{4}. It corresponds to
the replacement of the term $\ln\sigma(x,x')$ by the term $\ln
L^{-2}\sigma(x,x')$. Using the indeterminacy in the function
$F(x,x')$ we consider the replacement \cite{6}
\begin{equation}
    F(x,x')\longrightarrow F(x,x')+\frac{1}{4\pi}\ln L^2,
\end{equation}
where $L$ is a constant parameter with the dimension of length. In
quantum field theory it might find expression as an arbitrary
renormalization length or the Planck length. It is obvious that
this transformation do not change the constraints imposed on the
state dependent part of the two-pint function.

It is instructive to compare $\Sigma_{\alpha\beta}$ with the
renormalized stress tensor $\hspace {1cm}<T_{\alpha\beta}>_{ren}$.
The standard point-splitting renormalization method \cite{2a,3}
defines $<T_{\alpha\beta}>_{ren}$ as the limit
\begin{equation}\label{6a}
<T_{\alpha\beta}>_{ren}=\lim _{\hspace{-0.07cm}x\longrightarrow
x'}\frac{1}{2}{\mathcal{D}}_{\alpha\beta}(x,x')[G^{+}(x,x')-G_{ref}(x,x')]+\frac{1}{48\pi}R,
\end{equation}
where the differential operator
${\mathcal{D}}_{\alpha\beta}(x,x')$ is given by
\begin{equation}\label{6b}
{\mathcal{D}}_{\alpha\beta}(x,x')=\delta_\beta^{\hspace{0.2cm}\beta'}\nabla_\alpha\nabla_{\beta'}
-\frac{1}{2}g_{\alpha\beta}\delta_\rho^{\hspace{0.2cm} \rho'}
\nabla^\rho\nabla_{\rho'}
\end{equation}
and $\delta_\beta^{\hspace{0.2cm}\beta'}$ is the bitensor of
parallel displacement. The $G_{ref}$ is a reference two-point
function introduced in order to remove the singularities from
$G^+$. The second term in (\ref{6a}) is added to ensure the
conservation of the renormalized stress tensor. After some
manipulation one can show that the resulting expression of
(\ref{6a}) is equal to the $\Sigma_{\alpha\beta}$. According to
the uniqueness proof given in \cite{2a}, the tensor
$<T_{\alpha\beta}>_{ren}$ can be determined uniquely up to
addition of a conserved local curvature tensor, thus we have
\begin{equation}\label{7c}
<T_{\alpha\beta}(x)>_{ren}=\Sigma_{\alpha\beta}+\Gamma_{\alpha\beta},
\end{equation}
where $\Gamma_{\alpha\beta}$ is a state independent conserved
tensor which in a massless theory can only depend on the local
geometry. Following Wald's argument \cite{3} one can show that the
only geometrical conserved tensors in two dimensions, are those
obtained from a lagrangian of dimension (length)$^{-2}$. In a two
dimensional massless theory, one can only consider
${\mathcal{L}}=R$ which defines the vanishing conserved tensor
$$\Gamma_{\alpha\beta}=\frac{1}{\sqrt{-g}}\frac{\delta}{\delta g^{\alpha\beta}}\int d^2x
 \sqrt{-g}R=R_{\alpha\beta}-\frac{1}{2}g_{\alpha\beta}R.$$
 Therefore, we may take
$\Sigma_{\alpha\beta}$ as the quantum stress tensor induced by the
two-point function.
\section{Vacuum structure}

To characterize a Hadamard vacuum state, it is very essential to
determine $\Sigma_{\alpha\beta}$ corresponding to the vacuum
state. To see how $\Sigma_{\alpha\beta}$ contributes to the
Green's function $G^+(x,x')$, one can combine equations (\ref{6})
and (\ref{7}) to obtain
\begin{equation}
\begin{array}{ll}\label{8}\vspace{0.5cm}
F(x,x')=F(x)-\frac{1}{2}F_{;\alpha}\sigma^{;\alpha} +\frac{1}{2} [
\frac{1}{2}(F_{;\alpha\beta}(x)-\frac{1}{2}g_{\alpha\beta}\Box
F(x))\\ \hspace{1.5cm}
    -(\Sigma_{\alpha\beta}(x)+\frac{1}{48\pi}g_{\alpha\beta}R)]\sigma^{;\alpha}\sigma^{;\beta}+
    {\mathcal{O}}(\sigma^{3/2})\nonumber.
\end{array}
\end{equation}
It is obvious that any assumption about $\Sigma_{\alpha\beta}$
which respects the constraints (\ref{5}) and (\ref{5a}) for the
background metric in addition to a specific assumption on the
configuration of $F(x)$ can act as a superselection rule selecting
a local vacuum state and the corresponding Hilbert space. In the
following we proceed to present a dynamical model to determine the
configuration of the stress tensor $\Sigma_{\alpha\beta}$. we
choose  \cite{5a}
\begin{equation}\label{m1}
    F(x)= \psi(x),
\end{equation}
where $\psi(x)$ is taken to be a scalar field coupled to a two
dimensional gravitational background, with the action functional
\begin{equation}\label{m5}
    S=\int d^2x \sqrt{-g}(\frac{1}{2}\nabla^\alpha\psi\nabla_\alpha\psi+\zeta
    R\psi).
\end{equation}
This leads to the dynamical equation
\begin{equation}\label{m2}
    \Box\psi(x)-\zeta R=0,
\end{equation}
here $\zeta$ is a dimensionless constant. The basic assumption is
to relate the quantum stress tensor $\Sigma_{\alpha\beta}$ to the
stress tensor of the scalar field $\psi(x)$ \cite{5a}. We apply a
superselection rule of the form
\begin{equation}\label{m3}
\Sigma_{\alpha\beta}=T_{\alpha\beta}[\psi],
\end{equation}
where
\begin{equation}\label{m4}
    T_{\alpha\beta}=\frac{1}{2}(\psi_{;\alpha}\psi_{;\beta}-\frac{1}{2}g_{\alpha\beta}\psi^{;\rho}\psi_{;\rho})+
    \zeta g_{\alpha\beta}\Box\psi-\zeta\psi_{;\alpha;\beta},
\end{equation}
which can be obtained by varying the gravitational action
(\ref{m5}) with respect to $g^{\alpha\beta}$. The meaning of the
relation (\ref{m3}) is that it defines a formal prescription which
allows us to relate the tensor $F_{\alpha\beta}$ in (\ref{6}) to
the function $F(x)$ (or alternatively $\psi$) and the metric
tensor $g_{\alpha\beta}$. Thus, it characterizes a criterion to
select the admissible Hadamard vacuum states. The superselection
rule (\ref{m3}) should respects the constraints (\ref{5}) and
(\ref{5a}). The constraint
 (\ref{5}) is automatically satisfied through
(\ref{m2}). Satisfying the constraint (\ref{5a}) implies that
$\zeta^2=\frac{1}{24\pi}$. The conditions (\ref{m1}) and
(\ref{m3}) represent a vacuum structure in which the construction
of a local vacuum state is basically connected with the
determination of the scalar field $\psi$ through the solving
equation (\ref{m2}). On the other hand, the physical
characteristic of the solution of equation (\ref{m2}), depends
essentially on the boundary condition imposed on $\psi$.
Therefore, the choice of a boundary condition has an important
role to construct a vacuum state through this vacuum structure. In
the subsequent two sections we shall study this vacuum structure
using two different background metrics with different physical
characteristics.

\section{Hawking radiation}

In this section, we consider the two dimensional analog of the
Schwarzschild metric as the background metric
\begin{equation}
ds^2=-(1-\frac{2M}{r})dt^2+(1-\frac{2M}{r})^{-1}dr^2.
\end{equation}
We intend to study the physical characteristics of the quantum
stress tensor $\Sigma_{\alpha\beta}$ at sufficiently large
space-like distances $(r\rightarrow \infty)$. This consideration
can provide some information about the Hadamard vacuum state
corresponding to $\Sigma_{\alpha\beta}$. The dynamical equation
(\ref{m2}), at space-like infinity, reduce to
\begin{equation}\label{18}
    \Box\psi=0.
\end{equation}
We choose those solutions of (\ref{18}) which at sufficiently
 large distances are a function of the retarded time $t-r^*$,
 namely

\begin{equation}\label{19}
\psi(t,r)=U(t-r^*),\hspace{1cm}r^*=r+2M\ln(\frac{r}{2M}-1),
\end{equation}
here $U$ is an arbitrary function of the retarded time. Applying
this boundary condition on $\psi$ yields the configuration of the
quantum stress tensor
 at space-like infinity in the form
\begin{equation}\label{20}
 \Sigma^\alpha_\beta(r\rightarrow \infty)=(\frac{1}{2}\dot{U}^2-\frac{1}{\sqrt{24\pi}}\ddot{U})\left(%
\begin{array}{cc}
  -1 & -1 \\
  1 & 1 \\
\end{array}%
\right).
\end{equation}
It shows that the superselection rule implied by (\ref{m3}) leads
to a non-vanishing quantum stress tensor at space-like infinity.
Comparing (\ref{20}) with the stress tensor of a thermal radiation
\begin{equation}\label{20a}
    \frac{\pi}{12}(k_BT)^2\left(%
\begin{array}{cc}
  -1 & -1 \\
  1 & 1 \\
\end{array}%
\right),
\end{equation}
we infer that $\Sigma _{\alpha\beta}$ can describe a asymptotic
thermal radiation through the choice of the function $U(t-r^*)$ as
a solution of the equation
\begin{equation}\label{20b}
\frac{1}{2}\dot{U}^2-\frac{1}{\sqrt{24\pi}}\ddot{U}=\alpha^{-2},
\end{equation}
in which $\alpha$ is a constant parameter with the dimension of
length. The determination of $\alpha$ depends on the physical
characteristics of the state at hand. If $\alpha^2=768\pi M^2$,
one obtains an outward flux of radiation with the temperature
corresponding to the Hawking temperature $T=\frac{1}{8\pi}(k_B
M)^{-1}$ \cite{5}. This special choice for $\alpha$ can determine
a given vacuum state corresponding to the Hawking radiation.

\section{Cosmological model}
In this section, we investigate the physical characteristics of
the quantum stress tensor $\Sigma _{\alpha\beta}$ in a two
dimensional cosmological background described by the metric
\begin{equation}\label{m6}
    ds^2=-dt^2+a^2(t)dx^2.
\end{equation}
This is a two dimensional analog of the spatially flat
Friedman-Robertson-Walker space time. Here $a(t)$ is the scale
factor which we assume to follow a power law expansion
$a(t)=a_0(\frac{t}{t_0})^n $, with $t_0$ being the present age of
universe. The general solution of equation (\ref{m2}) in this
background metric, is
\begin{equation}\label{m7}
    \psi(t)=\frac{\gamma}{1-n}(\frac{t}{t_c})^{1-n}-\frac{2n}{\sqrt{24\pi}}\ln
    \frac{t}{t_c},
\end{equation}
here $\gamma$ is a dimensionless integration constant and $t_c$ is
an integration constant with the dimension of length. One can
interpret it as a cut-off time scale which is introduced in order
to exclude in the configuration of $\psi$ the contribution of the
early time singularity\footnote{Taking $t_c=0$ leads to a singular
behavior of $\psi$.}. This choice of the integration constants can
specify a given Hadamard vacuum state. Using (\ref{m7}) one can
obtain the non-vanishing components of the quantum stress tensor
$\Sigma_{\alpha\beta}$ through (\ref{m3}) in the form
\begin{equation}\label{m8}
      \Sigma^0_0=-\frac{\gamma^2}{4}(\frac{t_c}{t})^{2(n-1)}t^{-2}+\frac{n^2}{24\pi}t^{-2},
\end{equation}

\begin{equation}\label{m9}
     \Sigma^1_1=\frac{\gamma^2}{4}(\frac{t_c}{t})^{2(n-1)}t^{-2}+\frac{n(n-2)}{24\pi}t^{-2}.
\end{equation}
We intend to study the characteristics of the quantum stress
tensor at the present epoch. In this case, the expression of
$\Sigma_{\alpha\beta}$ coming from the variation of $\psi$ is
\begin{equation}\label{m10}
    \Sigma^\alpha_\beta(t\rightarrow t_0)=\frac{\gamma^2}{4}(\frac{t_c}{t_0})^{2(n-1)}t_0^{-2}\left(%
\begin{array}{cc}
  -1 & 0 \\
  0 & 1 \\
\end{array}%
\right)+{\mathcal{O}}(t_o^{-2}).
\end{equation}
Here the cut-off time $t_c$ is much smaller than $t_0$ and
$(\frac{t_c}{t_0})^{2(n-1)}\gg1$ for $n<1$ (The case $n<1$
corresponds to many cosmological models of physical interest
appropriate to a flat universe). Thus, in comparison with the
stress tensor of an equilibrium gas, namely
\begin{equation}\label{m11}
    \frac{\pi}{6}(k_BT)^2\left(%
\begin{array}{cc}
  -1 & 0 \\
  0 & 1\\
\end{array}%
\right),
\end{equation}
one may conclude that $\Sigma_{\alpha\beta}$ can approximately
describe the stress tensor of an equilibrium gas in the present
epoch, with the constant temperature
\begin{equation}\label{m12}
    T=\sqrt{\frac{3}{2\pi}}\gamma(\frac{t_c}{t_0})^{n-1}(k_Bt_0)^{-1}.
\end{equation}
 As one can see, these considerations can provide a non-vanishing vacuum energy at
the present epoch.
\section{Summary}
We have analyzed an analog of the Hadamard ansatz in two
dimensions  for the specification of the local Hadamard states
associated with a linear quantum field coupled to a two
dimensional gravitational background. To characterize a physical
state of interest, a superselection rule was applied in which the
state dependent part of the two-point function was related to a
dynamical scalar field. We applied this model using two different
background metrics and studied the characteristics of a vacuum
state through specifying the configuration of the quantum stress
tensor. These considerations provided the Hawking radiation at
space-like infinity, in the case of the Schwarzschild metric; and
a thermal radiation at the present epoch, in the context of
cosmology.

\end{document}